\documentstyle[12pt]{article}
 \makeatletter
 \def\leqq{\mathrel{\mathpalette\gl@align<}}
 \def\geqq{\mathrel{\mathpalette\gl@align>}}
 \def\gl@align#1#2{\lower.6ex\vbox{\baselineskip\z@skip\lineskip\z@
     \ialign{$\m@th#1\hfil##\hfil$\crcr#2\crcr=\crcr}}}
 \makeatother

 \makeatletter
 \def\sileqq{\mathrel{\mathpalette\gs@align<}}
 \def\sigeqq{\mathrel{\mathpalette\gs@align>}}
 \def\gs@align#1#2{\lower.6ex\vbox{\baselineskip\z@skip\lineskip\z@
     \ialign{$\m@th#1\hfil##\hfil$\crcr#2\crcr\sim\crcr}}}
 \makeatother
\begin{document}
\hbadness=10000
\hbadness=10000
\begin{titlepage}
\nopagebreak
\begin{flushright}
\end{flushright}
\vspace{1.5cm}
\begin{center}

{\large \bf Structure of Cubic Matrix Mechanics}

\vspace{1cm}
 
{Yoshiharu Kawamura}\footnote{E-mail:
haru@azusa.shinshu-u.ac.jp} 

\vspace{0.7cm}
Department of Physics, Shinshu University,
Matsumoto 390-8621, Japan\\
\end{center}
\vspace{0.7cm}

\nopagebreak

\begin{abstract}
We study the structure of cubic matrix mechanics
based on three-index objects.
It is shown that there exists a counterpart of canonical structure in
classical mechanics.
\end{abstract}
\vfill
\end{titlepage}
\pagestyle{plain}
\newpage
\def\thefootnote{\fnsymbol{footnote}}

\section{Introduction}

The study of the basic structure of mechanics is especially significant for several reasons.
First, such a study provides a unified description applicable to various systems.
For example, the canonical formalism in classical mechanics (CM) has been applied
to a wide range of dynamical systems,
from a system of point particles to continuous systems, such as relativistic field theories.\cite{CM}
Second it is useful in making clear the meaning of physical quantities
and symmetries of the system.
Third it is informative for the purpose of constructing 
a new mechanics based on the structure of established mechanics.
In fact, quantum mechanics (QM) was constructed with the aid of 
experimental results 
and the correspondence principle.\cite{CM-QM}

If there is a new mechanics beyond QM, it is natural to suppose that
it must possess a structure similar to that of QM and CM,
because a new scheme would contain QM as a limiting case, 
and the correspondence principle relates QM to CM.
Based on this conjecture, we consider the following questions:
\begin{enumerate}
\item What is the basic structure of such a new mechanics?

\item Is there a new, generalized mechanics beyond QM and CM?
\end{enumerate}

The first question can be rephrased as a question of
what structure of QM and CM is preserved and how other parts are modified.
We expect that this new scheme would contain descriptions of
time development and symmetry transformation properties of a system 
that are similar to those of QM and CM.
therefore, we require that {\it{the algebraic structure of equations of motion and symmetry transformations
be preserved (up to anomalous breakings) in the new mechanics}}.
Regarding the second question, a generalization of Heisenberg's matrix mechanics 
based on many-index objects has been proposed.\cite{YK}
It is important to explore the entire structure in order to obtain information on
the physical meaning and relation to reality. 

In this paper, we study the basic structure of cubic matrix mechanics
and show that this mechanics possesses a counterpart to the canonical structure in CM.

This paper is organized as follows. 
In the next section, we review the canonical structure of classical mechanics
and discuss the basic structure that a mechanics beyond CM should possess.
The structure of cubic matrix mechanics is studied in $\S$3. 
Section 4 is devoted to conclusions and discussion.

\section{Classical mechanics and beyond}

\subsection{Canonical structure of classical mechanics}

Here we review the canonical structure of classical mechanics.\cite{CM}
The canonical variables $q_a = q_a(t)$ and $p_a = p_a(t)$ satisfy Hamilton's canonical equations
\begin{eqnarray}
{d q_a \over dt} = {\partial H \over \partial p_a}, ~~
{d p_a \over dt} = - {\partial H \over \partial q_a} ,
\label{Hamilton-eq}
\end{eqnarray}
where $a= 1, \cdots ,n$ and $H$ is the Hamiltonian.
Physical variables are given by functions of the canonical variables and the time variable $t$; e.g., 
$A = A(q_a, p_a, t)$ and $B = B(q_a, p_a, t)$.
Hereafter we consider systems such that physical variables do not contain $t$ explicitly,
that is, closed physical systems.
The Poisson bracket of two variables $A$ and $B$ with respect to $q_a$ and $p_a$ 
is defined by
\begin{eqnarray}
\{A, B\}_{\scriptsize{\mbox{PB}}} \equiv \sum_{a =1}^n \left({\partial A \over \partial q_a}{\partial B \over \partial p_a}
- {\partial A \over \partial p_a}{\partial B \over \partial q_a}\right) .
\label{PB}
\end{eqnarray}
Hence, the Poisson brackets of the canonical variables are given by
\begin{eqnarray}
\{q_a, q_b\}_{\scriptsize{\mbox{PB}}} = \{p_a, p_b\}_{\scriptsize{\mbox{PB}}} = 0 , ~~ 
\{q_a, p_b\}_{\scriptsize{\mbox{PB}}} = \delta_{ab} .
\label{PB-qp}
\end{eqnarray}
The basic features of the Poisson bracket are as follows:
\begin{eqnarray}
&~& \{A, B\}_{\scriptsize{\mbox{PB}}} = - \{B, A\}_{\scriptsize{\mbox{PB}}} ,
~~~~~~~~~~~~~~~~~~~~~~~~~~~~{\mbox{(antisymmetry)}}
\label{PB-1}\\
&~& \{A + B, C\}_{\scriptsize{\mbox{PB}}} = \{A, C\}_{\scriptsize{\mbox{PB}}} 
+ \{B, C\}_{\scriptsize{\mbox{PB}}} ,~~~~~~~~~~~~{\mbox{(linearity)}}
\label{PB-2}\\
&~& \{\{A, B\}_{\scriptsize{\mbox{PB}}}, C\}_{\scriptsize{\mbox{PB}}} 
+ \{\{B, C\}_{\scriptsize{\mbox{PB}}}, A\}_{\scriptsize{\mbox{PB}}} 
+ \{\{C, A\}_{\scriptsize{\mbox{PB}}}, B\}_{\scriptsize{\mbox{PB}}} = 0 ,
\label{PB-3}\\
&~& \{A B, C\}_{\scriptsize{\mbox{PB}}} = A \{B, C\}_{\scriptsize{\mbox{PB}}} 
+ \{A, C\}_{\scriptsize{\mbox{PB}}} B . ~~~~~~~{\mbox{(derivation rule)}}
\label{PB-4}
\end{eqnarray}
By use of (\ref{Hamilton-eq}) and (\ref{PB}), the physical variable $A$ is shown to satisfy the equation
\begin{eqnarray}
{d A \over dt} = \{A, H\}_{\scriptsize{\mbox{PB}}} .
\label{Hamilton-eqA}
\end{eqnarray}

A transformation $A \to A' = {\cal{C}}(A)$ that preserves the Poisson bracket structure 
is called $\lq$canonical':
\begin{eqnarray}
\{A, B\}_{\scriptsize{\mbox{PB}}} \longrightarrow {\cal{C}}(\{A, B\}_{\scriptsize{\mbox{PB}}}) 
= \{{\cal{C}}(A), {\cal{C}}(B)\}_{\scriptsize{\mbox{PB}}} .
\label{caninical}
\end{eqnarray}
The infinitesimal version $A \to A' = A + \delta A$ is given by
\begin{eqnarray}
\delta A = \{A, G\}_{\scriptsize{\mbox{PB}}} \delta s ,
\label{inf-canonical-tr}
\end{eqnarray}
where $G$ is the generator of the transformation
and $\delta s$ is an infinitesimal parameter.
We can show that the Poisson bracket structure is preserved under 
the transformation (\ref{inf-canonical-tr}) by using the Jacobi identity (\ref{PB-3}).
When ${\cal{C}}(H) = H$, the transformation is a symmetry transformation,
and its generator $G$ is a constant of motion, i.e., $dG/dt = \{G, H\}_{\scriptsize{\mbox{PB}}} = 0$.
Then the equation of motion is form-invariant under the canonical transformation, so that
${d {\cal{C}}(A) / dt} = \{{\cal{C}}(A), {\cal{C}}(H)\}_{\scriptsize{\mbox{PB}}}$.
If $G_i$ are conserved quantities, $\{G_i, G_j\}_{\scriptsize{\mbox{PB}}}$ are also conserved,
as seen from (\ref{PB-3}) and (\ref{Hamilton-eqA}).

\subsection{Beyond classical mechanics}

The structure of classical mechanics is so simple and elegant that we expect that
a mechanics beyond CM must have a similar structure.
In this subsection, we present a conjecture for the basic structure of a new mechanics
by analogy to the canonical structure of CM.\footnote{
Though QM has been established as a realistic mechanics beyond CM,
we temporarily set QM aside to give full freedom to our imagination 
in the construction of a new scheme in this subsection.}

First we require the new scheme to possess the following properties:
\begin{enumerate}
\item There are counterparts of the canonical variables in CM, 
which are denoted $Q_a = Q_a(t)$ and $P_a = P_a(t)$, 
 and physical variables are functions
of $Q_a$ and $P_a$ in a closed system.\footnote{
In this paper, we do not consider a system with extented phase space variables,
described by Nambu.\cite{Nambu}
It is an interesting subject to study 
the relation between Nambu mechanics and generalized matrices.\cite{YK2}}
There exists a counterpart of the Poisson bracket, which we call $\lq$generalized bracket', and 
the bracket relations for $Q_a$ and $P_a$ are conditions that place restrictions on the phase space.
The generalized bracket does not necessarily possess all the algebraic properties 
of the Poisson bracket.
However, at least they possess properties of antisymmetry and linearity.

\item The same type of equation of motion holds for physical variables.
More specifically, the equation of motion is obtained by the replacement of the Poisson bracket 
with the generalized bracket.

\item There is a transformation that preserves the generalized bracket structure
that we call $\lq$a generalized canonical transformation'.
The Jacobi identity for the generalized bracket holds if it contains a conserved quantity.
Continuous symmetry transformations are realized as generalized canonical transformations 
generated by conserved quantities.
\end{enumerate}

Next, we formulate the basic structure of a new mechanics based on the above requirements.
\begin{enumerate}
\item Let us denote the generalized bracket by ${\cal{B}}(*, *)$ and impose the following conditions 
on $Q_a$ and $P_a$:
\begin{eqnarray}
{\cal{B}}(Q_a, Q_b) = {\cal{B}}(P_a, P_b) = 0 , ~~ {\cal{B}}(Q_a, P_b) = \delta_{ab} \Theta ,
\label{calB-qp}
\end{eqnarray}
where $\Theta$ is a constant of motion, and the bracket of $\Theta$ and a conserved quantity $\Lambda$ 
vanishes, i.e. ${\cal{B}}(\Theta, \Lambda) = 0$.
The antisymmetry and linearity conditions are expressed by
\begin{eqnarray}
&~& {\cal{B}}(A, B) = - {\cal{B}}(B, A) ,
\label{calB-1}\\
&~& {\cal{B}}(A + B, C) = {\cal{B}}(A, C) + {\cal{B}}(B, C) .
\label{calB-2}
\end{eqnarray}
We do not necessarily require the Jacobi identity 
nor the derivation rule as a property of ${\cal{B}}(*, *)$ for generic variables.

\item The equation of motion for a physical quantity $A$ is given by
\begin{eqnarray}
{d A \over dt} = {\cal{B}}(A, H) ,
\label{eq}
\end{eqnarray}
where $H$ corresponds to the Hamiltonian and is interpreted as the generator of time evolution
from the symmetry property discussed just below. 

\item A generalized canonical transformation is defined by the transformation $A \to A' = {\cal{G}}(A)$, 
which preserves the structure of ${\cal{B}}(*, *)$:
\begin{eqnarray}
{\cal{B}}(A, B) \longrightarrow {\cal{G}}({\cal{B}}(A, B))
= {\cal{B}}({\cal{G}}(A), {\cal{G}}(B)) . \label{g-canonical}
\end{eqnarray}
The infinitesimal version of (\ref{g-canonical}) is written
\begin{eqnarray}
\delta {\cal{B}}(A, B) = {\cal{B}}(\delta A, B) + {\cal{B}}(A, \delta B) ,
\label{inf-g-canonical}
\end{eqnarray}
under the infinitesimal generalized canonical transformation $A \to A' = A + \delta A$.
For a conserved quantity $G$, i.e., $dG/dt = {\cal{B}}(G, H) = 0$, 
the following Jacobi identity holds:
\begin{eqnarray}
{\cal{B}}({\cal{B}}(A, B), G) + {\cal{B}}({\cal{B}}(B, G), A) + {\cal{B}}({\cal{B}}(G, A), B) = 0 .
\label{calB-3}
\end{eqnarray}
Then, a symmetry transformation is given by 
the infinitesimal generalized canonical transformation,\footnote{
It is not certain whether every continuous generalized canonical transformation $A \to A' = {\cal{G}}(A)$
can be constructed from the infinitesimal one given by $\delta A = {\cal{B}}(A, F) \delta s$,
where $F$ is a generator.
Here, we require the algebraic structure of symmetry transformations to be identical to that in CM.}
\begin{eqnarray}
\delta A = {\cal{B}}(A, G) \delta s .
\label{inf-g-canonical-tr}
\end{eqnarray}
For conserved quantities $G_i$, ${\cal{B}}(G_i, G_j)$ are also conserved quantities,
as seen from (\ref{eq}) and (\ref{calB-3}).
\end{enumerate}

\subsection{Canonical formalism of quantum mechanics}
 
It is well known that there is a correspondence between the algebraic structure of quantum mechanics 
and that of classical mechanics.\cite{QM}
Here we give a brief review of the canonical formalism of quantum mechanics
using Heisenberg's matrix description for later convenience.
Physical quantities are represented by hermitian matrices that
can be written
\begin{eqnarray}
A_{mn}(t) = A_{mn} e^{i\Omega_{mn}t} = A_{mn} e^{{i \over \hbar}(E_m -E_n)t} ,
\label{Fmn} \end{eqnarray}
where the phase factor implies that a change in energy $E_{m} - E_n$ appears as 
radiation with angular frequency $\Omega_{mn}$, and the hermiticity of $A_{mn}(t)$
is expressed by $A_{nm}^{*}(t) = A_{mn}(t)$.
The quantity $A_{mn}(t)$ is a function of canonical pairs $(Q_a)_{mn}(t)$ and $(P_a)_{mn}(t)$.
A generalized bracket is a commutator divided by $i \hbar$:
\begin{eqnarray}
{\cal{B}}(A, B)_{mn} = {1 \over i\hbar} \sum_k (A_{mk}(t) B_{kn}(t) - B_{mk}(t) A_{kn}(t)) \equiv
{1 \over i \hbar}[A(t), B(t)]_{mn} .
\label{comm}
\end{eqnarray}
The commutators for the canonical pairs are given by
\begin{eqnarray}
[Q_a(t), Q_b(t)]_{mn} = [P_a(t), P_b(t)]_{mn} = 0 , ~~ [Q_a(t), P_b(t)]_{mn} = i \hbar \delta_{ab} \delta_{mn} .
\label{comm-QP}
\end{eqnarray}
It is obvious that the generalized bracket (\ref{comm}) possesses every property, 
including the Jacobi identity and the derivation rule, possessed by the Poisson bracket.


A physical quantity $A_{mn}(t)$ satisfies the Heisenberg equation
\begin{eqnarray}
{d \over dt}A_{mn}(t)  = {i \over \hbar}(E_m - E_n)A_{mn}(t)
= {1 \over i\hbar} [A(t), H]_{mn} ,
\label{Heisenberg-eq}
\end{eqnarray} 
where the Hamiltonian $H$ is a diagonal matrix written $H_{mn} \equiv E_m \delta_{mn}$.

A generalized canonical transformation is a unitary transformation given by 
\begin{eqnarray}
A_{mn}(t) \rightarrow A'_{mn}(t) = (U^{\dagger} A U)_{mn}(t) ,
\label{unitary-tr} \end{eqnarray}
where $U_{mn}$ is a unitary matrix.
The infinitesimal version is given by
\begin{eqnarray} \delta A_{mn}(t) = {1 \over i\hbar}[A(t), G]_{mn} \delta s .
\label{inf-unitary-tr}
\end{eqnarray}
For a unitary transformation generated 
by conserved quantities $(G_i)_{mn}$, 
$[G_i, G_j]_{mn}$ are also conserved.

\section{Cubic matrix mechanics}

We have discussed the basic structure that a new mechanics beyond classical mechanics should possess.
Quantum mechanics is a typical example, and it describes the microscopic world very successfully, 
but there is no definite reason 
that QM is the unique mechanics capable of describing nature at a fundamental level 
(around and beyond the gravitational scale).
For this reason, it is still meaningful to construct a new, generalized mechanics and study its properties.
In this section, we study the structure of cubic matrix mechanics, which has been recently proposed.\cite{YK}

\subsection{Cubic matrix}

Here we state our definition of a cubic matrix and its related terminology.
A cubic matrix is an object with three indices, $A_{lmn}$, 
which is a generalization
of a usual matrix, such as $B_{mn}$.\footnote{
Awata, Li, Minic and Yoneya introduced many-index objects
to quantize Nambu brackets.\cite{cubic}
We find that our definition of the triple product among cubic matrices 
is different from theirs, because we require a generalization of the Ritz rule
in the phase factor, but not necessarily the associativity of the products.}
We refer to a cubic matrix whose elements possess cyclic symmetry, i.e., $A_{lmn} = A_{mnl} = A_{nlm}$,
as a cyclic cubic matrix.
We define the hermiticity of a cubic matrix by 
$A_{l'm'n'}(t) = A_{lmn}^{*}(t)$ for odd permutations among indices
and refer to a cubic matrix with hermicity as a hermitian cubic matrix.
A hermitian cubic matrix is a special type of cyclic cubic matrix, because there it obeys a relation
$A_{lmn} = A_{mln}^{*} = A_{mnl} = A_{nml}^{*} = A_{nlm} = A_{lnm}^{*}$.
We refer to the following form of a cubic matrix as a normal form or a normal cubic matrix:
\begin{eqnarray}
A^{(N)}_{lmn} = \delta_{lm} a_{mn} + \delta_{mn} a_{nl} + \delta_{nl} a_{lm}  .
\label{AN} 
\end{eqnarray}
A normal cubic matrix is also a special type of cyclic cubic matrix.
The elements of a cubic matrix are treated as $c$-numbers throughout this paper.

\subsection{Cubic matrix mechanics and its structure}

The basic ingredient of this mechanics is a cyclic cubic matrix given by
\begin{eqnarray}
A_{lmn}(t) = A_{lmn} e^{i\Omega_{lmn}t} ,
\label{C}
\end{eqnarray}
where 
the angular frequency $\Omega_{lmn}$ has the form
\begin{eqnarray}
\Omega_{lmn} = \omega_{lm} - \omega_{ln} + \omega_{mn} \equiv
(\delta \omega)_{lmn} , ~~ \omega_{ml} = -\omega_{lm} .
\label{Omegalmn}
\end{eqnarray}
We assume the generalization of Bohrs' frequency condition\footnote{
Here and hereafter we use the reduced Planck constant $\hbar = {h \over 2\pi}$ 
as the unit of action in
our cubic matrix mechanics for simplicity.}
\begin{eqnarray}
\Omega_{lmn} = {1 \over \hbar} (E_{lm} - E_{ln} + E_{mn}) ,
\label{quanta}
\end{eqnarray}
where $E_{lm} (= - E_{ml})$ are energy eigenvalues.
The angular frequencies $\Omega_{lmn}$ have the properties
\begin{eqnarray}
\Omega_{l'm'n'} = \mbox{sgn}(P) \Omega_{lmn} , ~~
(\delta \Omega)_{lmnk} \equiv \Omega_{lmn} - \Omega_{lmk} + \Omega_{lnk}
 - \Omega_{mnk} = 0 ,
\label{cycle}
\end{eqnarray}
where sgn($P$) is $+1$ and $-1$ for even and odd permutation among indices, respectively.
The operator $\delta$ is regarded as a coboundary operator 
that changes $k$-th antisymmetric objects into $(k+1)$-th objects,
and this operation is nilpotent, i.e. $\delta^2(*) =0$.\cite{hom}
The frequency $\Omega_{lmn}$ is regarded as a $3$-coboundary.

If we define the triple product among cubic matrices $A_{lmn}(t) = A_{lmn} e^{i\Omega_{lmn}t}$, 
$B_{lmn}(t) = B_{lmn} e^{i\Omega_{lmn}t}$ and $C_{lmn}(t) = C_{lmn} e^{i\Omega_{lmn}t}$ by
\begin{eqnarray}
(A(t)B(t)C(t))_{lmn} \equiv
\sum_k A_{lmk}(t) B_{lkn}(t) C_{kmn}(t) = (A B C)_{lmn} e^{i\Omega_{lmn}t} ,
\label{cubicproduct}
\end{eqnarray}
this product takes the same form as (\ref{C}) with the relation (\ref{cycle}), which is
a generalization of the Ritz rule.
We comment that the resultant three-index object $(A B C)_{lmn} e^{i\Omega_{lmn}t}$ 
does not always have cyclic symmetry, even if $A_{lmn}(t)$, $B_{lmn}(t)$ and $C_{lmn}(t)$
are cyclic cubic matrices.
Note that this product is, in general, neither commutative nor associative; that is,
$(ABC)_{lmn} \neq (BAC)_{lmn}$ and
$(AB(CDE))_{lmn} \neq (A(BCD)E)_{lmn} \neq ((ABC)DE)_{lmn}$.
The triple-commutator is defined by
\begin{eqnarray}
&~& [A(t), B(t), C(t)]_{lmn} \equiv (A(t)B(t)C(t) + B(t)C(t)A(t) 
+ C(t)A(t)B(t) \nonumber \\
&~& ~~~ - B(t)A(t)C(t) - A(t)C(t)B(t) - C(t)B(t)A(t))_{lmn} .
\label{T-comm}
\end{eqnarray}
If $A_{lmn}(t)$, $B_{lmn}(t)$ and $C_{lmn}(t)$ are hermitian matrices,
$i[A(t), B(t), C(t)]_{lmn}$ is also a hermitian cubic matrix.  
The generalized bracket is defined by use of the triple-commutator (\ref{T-comm}) as
\begin{eqnarray}
{\cal{B}}(A, B)_{lmn} &\equiv& {1 \over i \hbar}[A(t), I, B(t)]_{lmn} ,
\label{cubic-bracket}
\end{eqnarray}
where $I$ is a special type of normal cubic matrix given by
\begin{eqnarray}
I_{lmn} = \delta_{lm} (1 - \delta_{mn}) + \delta_{mn} (1 - \delta_{nl}) + \delta_{nl} (1 - \delta_{lm})  .
\label{I}
\end{eqnarray}
By definition, we find that the generalized bracket (\ref{cubic-bracket}) has the properties of
antisymmetry and linearity, as seen from the relations
\begin{eqnarray}
&~& [A(t), I, B(t)]_{lmn} = - [B(t), I, A(t)]_{lmn} ,
\label{tricomm-1}\\
&~& [A(t) + B(t), I, C(t)]_{lmn} = [A(t), I, C(t)]_{lmn} + [B(t), I, C(t)]_{lmn} .
\label{tricomm-2}
\end{eqnarray}
Note that neither the derivation rule nor the Jacobi identity necessarily holds for generic variables.
(See the appendix for features of the triple-commutator $[A, I, B]$.)

We impose the following conditions on the canonical pairs $(Q_a)_{lmn}$ and $(P_a)_{lmn}$:
\begin{eqnarray}
&~& [Q_a(t), I, Q_b(t)]_{lmn} = [P_a(t), I, P_b(t)]_{lmn} = 0 , \nonumber \\
&~& [Q_a(t), I, P_b(t)]_{lmn} = i \hbar \delta_{ab} \Theta_{lmn}.
\label{qc-QP}
\end{eqnarray}
Here $\Theta_{lmn}$ can be a normal cubic matrix,
because the conditions should be time-independent, and a normal cubic
matrix is a constant of motion, as seen below.

The cyclic cubic matrix $A_{lmn}(t)$ yields the generalization of 
the Heisenberg equation,
\begin{eqnarray}
{d \over dt}A_{lmn}(t) = i \Omega_{lmn} A_{lmn}(t) 
= {1 \over i\hbar} [A(t), I, H]_{lmn} ,
\label{cH-eq}
\end{eqnarray}
where $H$ is the Hamiltonian written
\begin{eqnarray}
H_{lmn} = {1 \over 2} \delta_{lm} E_{mn}
+ {1 \over 2} \delta_{mn} E_{nl} + {1 \over 2} \delta_{nl} E_{lm} .
\label{H} 
\end{eqnarray}
Because the Hamiltonian is a normal form, we find that an arbitrary normal cubic matrix $A^{(N)}$ 
is a constant of motion: $i \hbar (dA^{(N)}/dt)_{lmn} = [A^{(N)}, I, H]_{lmn} = 0$. 

The generalized bracket structure (\ref{cubic-bracket}) is preserved by the infinitesimal 
transformation
\begin{eqnarray}
\delta A_{lmn}(t) = {1 \over i\hbar}[A(t), I, G^{(N)}]_{lmn} \delta s ,
\label{c-inf-unitary-tr}
\end{eqnarray}
where $G^{(N)}$ is a normal cubic matrix, that is, a generator of the symmetry transformation.
Here, we use the fact that the Jacobi identity holds for $G^{(N)}$, so that
\begin{eqnarray}
[[A, I, B], I, G^{(N)}]_{lmn} + [[B, I, G^{(N)}], I, A]_{lmn} + [[G^{(N)}, I, A], I, B]_{lmn} = 0 .
\label{C-Jacobi}
\end{eqnarray}
We find that the derivation rule
\begin{eqnarray}
&~& [ABC, I, G^{(N)}]_{lmn} = (AB[C, I, G^{(N)}])_{lmn} + (A[B, I, G^{(N)}]C)_{lmn} \nonumber \\
&~& ~~~~~~~~~~~~~~~~~~~~~~~~~~~~~~~~~~~~~ + ([A, I, G^{(N)}]BC)_{lmn}  
\label{C-D-rule}
\end{eqnarray}
holds for $G^{(N)}$ if $(ABC)_{lmn}$ is a cyclic cubic matrix.

\section{Conclusions and discussion}

We have studied the basic structure of a cubic matrix mechanics
and shown that this mechanics meets the requirement 
that {\it{the algebraic structure of equations of motion and symmetry transformations
be preserved (up to anomalous breakings) in any new mechanics beyond quantum mechanics and 
classical mechanics.}}

The basic structure of this mechanics is summarized as follows.
There is a symmetry transformation for a physical quantity $A$ [which is a cyclic cubic matrix in the cubic
matrix mechanics], whose infinitesimal version 
is given by $\delta A = {\cal{B}}(A, G) \delta s$,
where ${\cal{B}}(*, *)$ is the counterpart of the Poisson bracket in CM
[${\cal{B}}(A, G) \equiv {1 \over i\hbar}[A, I, G]$ in cubic matrix mechanics], 
and $G$ is the generator of the transformation.
The time evolution of $A$ is regarded as a symmetry transformation generated by
the Hamilotonian $H$: $\delta A = {\cal{B}}(A, H) \delta t$
[which is the generalization of the Heisenberg equation (\ref{cH-eq}) in cubic matrix mechanics].
The generator $G$ is a constant of motion, i.e. $dG/dt = {\cal{B}}(G, H) = 0$.
[The normal form $G^{(N)}_{lmn}$ is a conserved quantity in cubic matrix mechanics].
The Jacobi identity holds 
in the case that it contains a conserved quantity such as (\ref{calB-3})
[(\ref{C-Jacobi}) in cubic matrix mechanics].
The bracket structure is preserved under the symmetry transformation generated by a normal cubic matrix
in cubic matrix mechanics,
as seen from the Jacobi identity.

The above-stated requirement stems from the expectation that a conservation law survives beyond QM,
but it may be too naive and need some modifications if we wish to incorporate a gravitational interaction.
In fact, there is the conjecture that no continuous global symmetry exists 
in a quantum theory including gravity.\cite{QG}
Moreover, the theory should be formulated in a background-independent way, as the theory of general relativity.
Therefore, the new scheme discussed in this paper can be interpreted as an effective description of an
underlying mechanics after fixing the background geometry and ignoring dynamical degrees of freedom
for the graviton.

There are interesting subjects that remain.
One is the construction of a matrix mechanics in which physical variables are $n$-index objects
($n \geq 4$).
This was proposed in Ref. \cite{YK}, but its formulation has not yet been completed.
A conjecture for $\lq\lq$Hamiltonians" and a generalized bracket has been given, but there is the limitation
that the generalized Heisenberg equation holds only 
for $n$-index objects whose indices are completely different.
Another subject is the study of the relation between cubic matrix mechanics and Heisenberg matrix mechanics.
Such studies are now in progress.

\section*{Acknowledgements}
We would like to thank Prof. S. Odake for useful discussions 
and a referee who gave several useful comments and pointed out some mistakes.

\appendix
\section{}

In this appendix, we study features of the triple-commutator $[A, I, B]$
for cyclic cubic matrices $A_{lmn}$ and $B_{lmn}$.
This commutator is written
\begin{eqnarray}
[A, I, B]_{lmn} = A_{lmn} \widetilde{B}_{lmn} - B_{lmn} \widetilde{A}_{lmn} + (AB)^0_{lmn} ,
\label{AIB}
\end{eqnarray}
where $\widetilde{B}_{lmn}$ and $(AB)^0_{lmn}$ are defined by
\begin{eqnarray}
\widetilde{B}_{lmn} \equiv B_{lmm} - B_{mll} + B_{mnn} - B_{nmm} + B_{nll} - B_{lnn} ~~~~~~~~~~~~~~~~~~~~~~~~
\label{tildeB} 
\end{eqnarray}
and 
\begin{eqnarray}
&~& (AB)^0_{lmn} \equiv \delta_{lm} \sum_{k} (A_{mnk} B_{nmk} - B_{mnk} A_{nmk})
  + \delta_{mn} \sum_{k} (A_{nlk} B_{lnk} - B_{nlk} A_{lnk}) \nonumber \\
&~& ~~~~~~~~~~~~~~~~~~~ + \delta_{nl} \sum_{k} (A_{lmk} B_{mlk} - B_{lmk} A_{mlk})  ,
\label{AB0}
\end{eqnarray}
respectively.
The features of $\widetilde{B}_{lmn}$ and $(AB)^0_{lmn}$ are as follows:
\begin{enumerate}
\item  $\widetilde{B}_{lmn}$ is skew-symmetric:
\begin{eqnarray}
\widetilde{B}_{lmn} = \widetilde{B}_{mnl} = \widetilde{B}_{nlm} 
= - \widetilde{B}_{nml} = - \widetilde{B}_{mln} = - \widetilde{B}_{lnm} .
\label{B-cyclic}
\end{eqnarray}


\item  $\widetilde{B}_{lmn}$ is a $\delta$-closed (3-cocycle): 
\begin{eqnarray}
(\delta{\widetilde{B}})_{lmnk} \equiv 
\widetilde{B}_{lmn} - \widetilde{B}_{lmk} + \widetilde{B}_{lnk} - \widetilde{B}_{mnk} = 0 .
\label{B-2b}
\end{eqnarray}

\item $(AB)^0_{lmn}$ is a normal cubic matrix with $(AB)^0_{lmn} = - (BA)^0_{lmn}$.

\end{enumerate}

For a normal cubic matrix $A^{(N)}_{lmn}= \delta_{lm} a_{mn} + \delta_{mn} a_{nl} + \delta_{nl} a_{lm}$, 
$\widetilde{{A}^{(N)}}_{lmn}$ is given by
\begin{eqnarray}
\widetilde{{A}^{(N)}}_{lmn} 
&=& {{A}^{(N)}}_{lmm} - {{A}^{(N)}}_{mll} + {{A}^{(N)}}_{mnn} 
  - {{A}^{(N)}}_{nmm} + {{A}^{(N)}}_{nll} - {{A}^{(N)}}_{lnn} \nonumber \\
&=& - 2(a^{(-)}_{lm} + a^{(-)}_{mn} + a^{(-)}_{nl}) ,
\label{tildeAN} 
\end{eqnarray}
where $a^{(-)}_{lm} \equiv {1 \over 2}(a_{lm} - a_{ml})$.

We obtain the following relations for cyclic cubic matrices 
from the above expressions and features.
\begin{enumerate}
\item For an arbitrary cyclic cubic matrix $A_{lmn}$ and an arbitrary normal cubic matrix $B^{(N)}_{lmn}$, 
the relations $(AB^{(N)})^0_{lmn} = 0$ and $[A, I, B^{(N)}]_{lmn} = A_{lmn}\widetilde{{B}^{(N)}}_{lmn}$
hold.

\item Arbitrary normal cubic matrices $A^{(N)}_{lmn}$ and $B^{(N)}_{lmn}$
 commute; 
\begin{eqnarray}
[A^{(N)}, I, B^{(N)}]_{lmn} = 0.
\label{identity}
\end{eqnarray}


\item The Jacobi identity holds if any of $A$, $B$ or $C$ has a normal form
seen from the relation
\begin{eqnarray}
&~& [[A, I, B], I, C]_{lmn} + [[B, I, C], I, A]_{lmn} + [[C, I, A], I, B]_{lmn} \nonumber \\
&~& ~~~~~ = -\widetilde{{({AB})}^0}_{lmn} C_{lmn} - \widetilde{{({BC})}^0}_{lmn} A_{lmn} 
- \widetilde{{({CA})}^0}_{lmn} B_{lmn} .
\label{Jacobi}
\end{eqnarray}

\item The derivation rule holds for arbitrary normal cubic matrices $D^{(N)}_{lmn}$, so that
\begin{eqnarray}
&~& [ABC, I, D^{(N)}]_{lmn} = (AB[C, I, D^{(N)}])_{lmn} + (A[B, I, D^{(N)}]C)_{lmn} \nonumber \\
&~& ~~~~~~~~~~~~~~~~~~~~~~~~~~~ + ([A, I, D^{(N)}]BC)_{lmn}  ,
\label{D-rule}
\end{eqnarray}
if $(ABC)_{lmn}$ is a cyclic cubic matrix.
Here we have used the fact that $\widetilde{{D}^{(N)}}_{lmn}$ is a 3-cocycle.
\end{enumerate}

\end{document}